\documentclass[pra,twocolumn,a4paper,showpacs,floatfix]{revtex4}
\usepackage{epsfig} 
\begin{document}
\title{Aharonov-Bohm ring with a side-coupled atomic cluster: magneto-transport
and the selective switching effect} 
\author{Supriya Jana and Arunava Chakrabarti}
\affiliation{Department of Physics \\ University of Kalyani \\
Kalyani \\ West Bengal 741 235, India} 
\begin{abstract}
\begin{center}
{\bf Abstract}
\end{center}
\vskip .25in
We report electronic transmission properties of a simple tight binding 
Aharonov-Bohm ring threaded by a magnetic flux, to one arm of 
which a finite cluster of atoms has been attached from one side. We demonstrate that, 
by suitably choosing the number of scatterers in each arm of the quantum ring, the
transmission across the ring can be completely blocked when the ring is decoupled from 
the atomic cluster and the flux threading the 
ring becomes equal to half the fundamental flux quantum. A novel transmission resonance
then occurs immediately as the coupling between the ring and the impurity cluster is
switched `on'. It is shown that the delta-like transmission resonances occur precisely at 
the eigenvalues of the side coupled chain of atoms. 
The `switching' effect can be observed either for all the eigenvalues of the 
isolated atomic cluster, or for a selected set of them, depending on the number of
scatterers in the arms of the ring.
The ring-dot coupling can be gradually increased to
suppress the oscillations in the magneto-transmission completely. However, the suppression can lead either 
to a complete transparency or no transmission at all, occasionally accompanied by a 
reversal of phase at special values of the magnetic flux. 
\end{abstract}
\maketitle 
\vskip .25in
\noindent
\section{Introduction}
\vskip .25in
Simple tight binding models of mesoscopic systems have been quite
extensively studied in recent times \cite{kub02}-\cite{xiong06} with a 
view to understand the basic features of electronic transport in quantum dots (QD) or
the magneto-transport in closed loop geometries such as an Aharonov-Bohm (AB) ring. 
One reason behind such model studies is definitely
the simple geometry of the models which enables one to derive exact results and 
to look into the possible causes of certain salient features observed in the transport 
properties of 
real life small scale semiconducting or metallic systems. The other reason can be attributed  
to the immense success of nano-technology and the use of precision instruments such
as a scanning tunnel microscope (STM) which can be used to build low dimensional
nanostructures with tailor-made geometries.
One such geometry, which will be our concern in this communication, is an AB ring 
threaded by a magnetic flux $\Phi$ and with 
a finite segment of $N$ atomic sites attached to one arm of the ring at an arbitrary point.

The central feature of electron transport across an AB ring is the periodic oscillation 
in the magnetoconductance whenever the phase coherence length exceeds the dimension of the sample 
\cite{ab59}. B\"{u}ttiker et al \cite{butt83} provided an early formulation of the problem. The 
transport in such a closed geometry was re-addressed by Gefen et al \cite{gefen84} who obtained an 
exact expression for the two-terminal conductance across the ring. Using a discrete tight binding 
formulation the two-terminal conductance was also examined by D'Amato et al \cite{amato89}, and 
subsequently, by Aldea et al \cite{aldea92}. A non-trivial change in the transport of an AB ring 
is observed when the ring contains a quantum dot (QD) either embedded in an arm, or side coupled 
to it \cite{yacob95}-\cite{holl01}. 
Motivated by the experiment of Yacoby et al \cite{yacob95}, Yeyati and B\"{u}ttiker \cite{levy95} 
prescribed an exact formulation of the magnetoconductance of an AB ring with a QD embedded in its arm.
A similar problem with a multi-terminal geometry was later addressed by Kang \cite{kang99}.
The QD-AB ring hybrid system also received attention in relatively recent experiments by 
Meier et al \cite{meier04} and Kobayashi et al \cite{koba02}-\cite{koba04} with a focus on 
the study of single electron charging and suppression of AB oscillations \cite{meier04}, and the Fano 
resonance in the magnetoconductance \cite{koba02}-\cite{koba04}.

In a quantum dot discrete energy levels arise as a
consequence of confinement of electrons in all three directions. This has inspired a considerable
number of theoretical works involving a discrete lattice of the so called `single level' QD's \cite{kub02}, 
mimicked by `atomic' sites arranged either 
in an open geometrical arrangement or in a closed AB ring within a tight binding formalism. 
For example, QD's, single, or in an array, side coupled to an open chain 
have already received attention in the context of Kondo effect \cite{torio02}, one electron transport 
\cite{ore03}, \cite{rod03}, \cite{adam04} and Dicke effect \cite{paore04}. The prospect of 
engineering Fano resonances \cite{mir05}-\cite{kiv05}, design of spin filters \cite{lee06}
and the localization-delocalization problem \cite{pou02}
in a series of atomic clusters side coupled to an infinite lattice have also been discussed in details. 

Motivated by such simple models which, inspite of their simplicity, brings out the rich 
quantum coherence effects exhibited by a mesoscopic system, we re-visit the problem 
of magneto-transport in an AB ring threaded by a magnetic flux, but now  
with a chain of $N$ atomic sites (an array of {\it single level QD's}) 
attached to one arm of the ring. Inspite of the previous studies
we believe that the interplay of the closed loop geometry and the eigenvalue spectrum 
of the dangling QD array is little studied and, is likely to provide new features in the electronic transport, 
that might throw some light on the potential of such systems as novel quantum devices.  
This is our main objective.

We focus on the role of the ring-dot coupling in particular, and 
come across several interesting results. For example, it is found that the ring-dot system 
displays a `switching' action for small values of the ring-dot coupling 
at selected energies of the electron when the flux penetrating the 
ring is $\Phi=\Phi_0/2$. The energy values (at which the switching takes place), for a given size 
of the QD array coupled to the ring, 
belong to the set of eigenvalues of the isolated QD array, and  
depend on the number of scatterers in the upper and the lower arms of the ring. 
A gradual increase in the ring-dot coupling suppresses the AB oscillations and 
is accompanied by occasional transmission phase-reversals at specific values of the 
magnetic flux.  

In what follows, we present the model and the method in section II. Section III contains the results and the related 
discussion and we draw conclusion in section IV.

\begin{center}
\begin{figure}
{\centering \resizebox* 
{6cm}{5cm}{\includegraphics[angle=0]{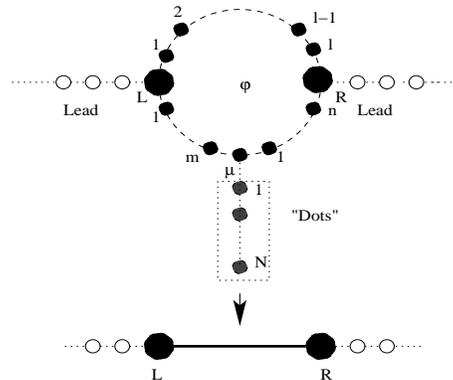}}}
\caption{\label{fig1}The quantum ring and the side coupled chain of quantum dots. 
The sites $L$ and $R$ (bigger solid circles) mark the left and the right junctions with the leads, and 
the site marked $\mu=+1$ is the `connecting' point of the side coupled FA chain.} 
\end{figure}
\end{center}
\vskip .25in
\noindent
\section{The model and the method}
\vskip .25in
We begin by referring to Fig.1. The ring contains $l$ atoms in the upper arm, excluding the 
sites marked $L$ and $R$ where the leads join the ring. The site marked $\mu$ in the lower arm 
is the point where the dangling QD chain  
is attached. There are $m$ sites to the left of $\mu$ and 
$n$ sites to its right. Changing $m$ and $n$ therefore shifts the location of the attachment 
of the `defect' chain. The Hamiltonian of the lead-ring-dot-lead system, in the standard tight 
binding form, is written as, 
\begin{equation}
H = H_{lead} + H_{ring} + H_{dot} + H_{ring-dot} + H_{ring-lead}
\end{equation}
where,  
\begin{eqnarray}
H_{lead} & = & \epsilon_0\sum_{i=-\infty}^{L-1} c_i^{\dag}c_i + 
\epsilon_0\sum_{i=R+1}^{\infty} c_i^{\dag}c_i + 
t_0 \sum_{<ij>} c_i^{\dag} c_j \nonumber \\
H_{ring} & = & \epsilon_L r_L^{\dag}r_L + \epsilon_R r_R^{\dag}r_R + 
t_0 \exp(i\gamma)\sum_{<ij>} r_i^{\dag}r_j + h.c. \nonumber \\
H_{dot} & = & \sum_{i=1}^{N} \epsilon_i d_i^{\dag} d_i + t_0  
\sum_{i=1}^{N-1} d_i^{\dag} d_j + h.c.\nonumber \\
H_{ring-dot} & = & \lambda (r_{\mu}^{\dag}d_1 + h.c.) \nonumber \\
H_{ring-lead} & = & t_0 (r_L^{\dag}c_{L-1} + r_R^{\dag}c_{R+1}) 
\end{eqnarray}
In the above, $c^{\dag}(c)$, $r^{\dag}(r)$, and $d^{\dag}(d)$ represent the 
creation (annihilation) operators for the leads, the ring and the QD chain 
respectively. $r_L$$(r_L^\dag)$ and $r_R(r_R^\dag)$ represent the same at the lead-ring connecting
 sites $L$ and $R$ respectively.
The on-site potential at the leads, in the QD chain  and in the bulk of the ring is taken to be  
 $\epsilon_0$ for every site including the site marked $\mu$. 
The lead-ring connecting sites have been assigned the 
on-site potentials $\epsilon_L$ and $\epsilon_R$ respectively. The amplitude of the hopping integral
is taken to be $t_0$ throughout except the hopping from the site $\mu$ in the ring to the first site
of the QD chain, which has been symbolized as $\lambda$ and represents the `strength' of coupling 
between the ring and the QD array. $\gamma$ is given by 
$\gamma = 2\pi \Phi/(l+m+n+1)\Phi_0$, where, $\Phi$ is the flux threading the ring and $\Phi_0=hc/e$ is 
the fundamental flux quantum.
The task of solving the Schr\"{o}dinger equation to obtain the 
stationary states of the system can be reduced to an equivalent problem of solving a set of 
the following difference equations:

For the sites $L$ and $R$ at the ring-lead junctions, 
\begin{eqnarray}
(E-\epsilon_L) \psi_L & = & t_0e^{i\gamma} \psi_{1,\cal U} + t_0e^{-i\gamma} \psi_{1,\cal L} + t_0 \psi_{L-1} \nonumber \\
(E-\epsilon_R) \psi_R & = & t_0e^{-i\gamma} \psi_{l,\cal U} + t_0e^{i\gamma} \psi_{m+n+1,\cal L} + t_0 \psi_{R+1} 
\nonumber \\
\end{eqnarray}
In the above, $\psi_{L-1}$ and $\psi_{R+1}$ represent the amplitudes of the wave function at the sites on the 
lead which are closest to the points $L$ and $R$, and,  
$\cal U$ and $\cal L$ in the subscripts refer to the `upper' and the `lower' arms respectively.

For the sites in the bulk of the ring the equations are,
\begin{eqnarray}
(E-\epsilon_0) \psi_{j,\cal U} & = & t_0e^{-i\gamma} \psi_{j-1,\cal U} + t_0e^{i\gamma} \psi_{j+1,\cal U} \nonumber \\
(E-\epsilon_0) \psi_{j,\cal L} & = & t_0e^{i\gamma} \psi_{j-1,\cal L} + t_0e^{-i\gamma} \psi_{j+1,\cal L} 
\end{eqnarray}
where, by $j+1$ and $j-1$th we symbolize the sites to the right and to the left of the $j$th site in any
arm of the ring.

For the site marked $\mu$ in the lower arm of the ring, the equation is, 
\begin{equation}
(E-\epsilon_0) \psi_{\mu,\cal L} = t_0 e^{i\gamma}\psi_{\mu-1,\cal L} + t_0 e^{-i\gamma}\psi_{\mu+1,\cal L}
\end{equation}
$\mu \pm 1$ implying the sites to the right and to the left of the site marked $\mu$ respectively. Finally, 
for the QD array we have the following set of difference equations,
\begin{eqnarray}
(E-\epsilon_1) \psi_1 & = & \lambda \psi_\mu + t_0 \psi_2 \nonumber \\
(E-\epsilon_j) \psi_j & = & t_0 \psi_{j-1} + t_0 \psi_{j+1} \nonumber \\
(E-\epsilon_N) \psi_N & = & t_0 \psi_{N-1}
\end{eqnarray}
where the central set of equations above refer to the bulk sites viz. $j=2,...N-1$ in the QD array.

The process of calculating the transmission coefficient across such a ring-dot 
system consists of the following steps. First, the dangling QD chain is `wrapped' into an effective 
site by decimating the amplitudes $\psi_2$ to $\psi_N$ from Eq.(6). 
The renormalized on-site potential of the first site of the QD array is given by \cite{arun07}, 
\begin{equation}
\tilde\epsilon = \epsilon_0 + 
\frac{t_0U_{N-3}(x)}{U_{N-2}(x)}+\frac{t_0^2}{U_{N-2}^2(x)}
\frac{1}{E-\epsilon_0-\frac{t_0U_{N-3}(x)}{U_{N-2}(x)}}
\end{equation}
for $N \ge 2$. For $N=1$, we simply have $\tilde\epsilon = \epsilon_0+\lambda^2/(E-\epsilon_0)$.
Here, $x=(E-\epsilon_0)/2t_0$ and $U_N(x)$ is the $N$th order Chebyshev 
polynomial of the second kind, with $U_0=1$ and $U_{-1}=0$ \cite{grim00}.
This `effective' site is coupled to the site marked $\mu$ in the lower arm of the ring via a 
hopping integral $\lambda$. In the second step, the effective site with on-site potential is further 
`folded' back into the site $\mu$, whose renormalized on-site potential now reads \cite{arun07}, 
\begin{equation}
\epsilon^* = \epsilon_0 + \frac{\lambda^2}{E-\tilde\epsilon}.
\end{equation}
We now have a ring with $l$ atoms in the upper arm and an effective site at position $\mu$ in the 
lower arm, flanked by $m$ atoms on its left and $n$ atoms on the right, so that there is a total 
of $m+n+1$ atoms in the lower arm. This is just the case of a QD with an energy dependent on-site 
potential embedded in an arm of an AB-ring.
In the final step, all the $l+m+n+1$ atoms are decimated using the set 
of appropriate difference equations (3) and (4), to reduce the ring into an effective 
{\it diatomic molecule} (Fig.1). 
The renormalized values of the on-site potential at the two extremeties of the molecule are given by, 
\begin{eqnarray}
\tilde\epsilon_L & = & \epsilon_0 + t_0 \left (\frac{U_{l-1}}{U_l} + \frac{U_{m-1}}{U_m} \right ) +
\frac{t_0^2}{U_m^2} F(E,\lambda,m,n) \nonumber \\
\tilde\epsilon_R & = & \epsilon_0 + t_0 \left (\frac{U_{l-1}}{U_l} + \frac{U_{n-1}}{U_n} \right ) +
\frac{t_0^2}{U_n^2} F(E,\lambda,m,n) \nonumber \\
\end{eqnarray}
where, for a fixed set of $\epsilon_0$ and $t_0$,
\begin{equation}
F(E,\lambda,m,n) = \left [E-\epsilon^*-t_0 \left (\frac{U_{m-1}}{U_m} + \frac{U_{n-1}}{U_n}\right )\right ]^{-1}
\end{equation}
The time reversal symmetry of the hopping integral between the atoms at $L$ and $R$ of the diatomic
 molecule is broken due to the flux threading the ring, and is given by, 
\begin{equation}
t_F = \frac{t_0}{U_l} e^{i(l+1)\gamma} + \frac{t_0^2}{U_mU_n} F(E,\lambda,m,n) e^{-i(m+n+2)\gamma}
\end{equation}
for the {\it forward} hopping from $L$ to $R$ and by $t_B=t_F^*$ for the {\it backward} hopping from $R$ to 
$L$.
The transmission coefficient across the effective diatomic molecule is given by \cite{stone81}, 
\begin{widetext}
\begin{equation}
T = \frac{4 \sin^2 qa}{|M_{12}-M_{21}+(M_{11}-M_{22}) \cos qa|^2 + |M_{11}+M_{22}|^2 \sin^2 qa}
\end{equation}
\end{widetext}
where, $a$ is the lattice constant in the leads, taken to be equal to one throughout the calculation.
In what follws, we discuss various aspects of the electronic transmission across the ring-QD array system.
We fix the on-site potential at all sites, including the QD chain, as $\epsilon_0$ and the 
hopping integrals has been kept equal to $t_0$ throughout, except the ring-QD array coupling $\lambda$.
The `defect' that we hang from an otherwise perfect ring is thus only of a topological nature.
\vskip .25in
\section{Results and Discussions}
\subsection{Suprression of AB oscillations}
\begin{center}
\begin{figure}
{\centering \resizebox* 
{16cm}{14cm}{\includegraphics[angle=-90]{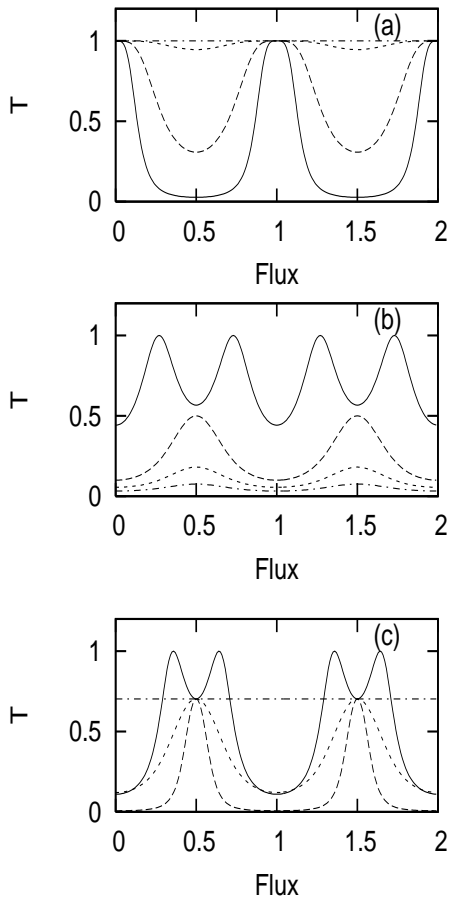}}}
\caption{\label{fig2}The gradual suppression of AB oscillations in the transmission
spectrum as the ring-dot coupling increases. (a) $E=10^{-6}$, $l=13$, $m=n=6$ and $N=5$. Here, 
$\lambda=0.001$ (solid), $0.002$ (dashed), $0.005$ (dotted) and $0.1$ (dot-dash). (b) $E=10^{-6}$, 
$L=7$, $m=3$, $n=8$ and $N=7$. $\lambda=0.001$ (solid), $0.004$ (dashed), $0.005$ (dotted) and $0.006$ (dot-dash).
(c) $E=0.618011988$, $l=9$, $m=3$, $n=8$ and $N=4$. The values of $\lambda$ are, $0.001$ (solid), $0.007$ (dashed), 
$0.01$ (dotted) and $0.5$ (dot-dash) respectively. Other parameters are, $\epsilon_0=0$, $t_0=1$ and the energy and $\lambda$
are measured in unit of $t_0$, and the flux is measured in unit of $\Phi_0$}
\end{figure}
\end{center}

In all transmission profiles the ring-dot coupling $\lambda$ plays a crucial role. The first 
effect that we present is a supression of the AB oscillations as a function of $\lambda$. 
We choose the energy $E$ from a  specially selected set obtained by solving the equation $E-\tilde\epsilon=0$. 
For these $E$-values the suppression of the 
 AB oscillations can be directly worked out from our formulation. It is to be appreciated that  
the eigenvalues of the isolated quantum dot array are obtained by solving the polynomial equation 
$E-\tilde\epsilon = 0$ \cite{mir05}-\cite{arun07}. We select any one of the roots, 
name it $\tilde\epsilon_0$, and fix $\lambda \ne 0$. This last condition is important.

A close look at the expression of $\epsilon^*$ reveals that for $E=\tilde\epsilon_0$ (in fact, for 
any real root of the equation $E-\tilde\epsilon = 0$) we get $\epsilon^*=\infty$. This leads to the 
following reduced forms of $\epsilon_L$, $\epsilon_R$ and $t_F (=t_B^*)$:
\begin{eqnarray}
\tilde\epsilon_L & = & \epsilon_0 + t_0 \left (\frac{U_{l-1}}{U_l} + \frac{U_{m-1}}{U_m} \right ) \nonumber \\
\tilde\epsilon_R & = & \epsilon_0 + t_0 \left (\frac{U_{l-1}}{U_l} + \frac{U_{n-1}}{U_n} \right ) \nonumber \\
t_F & = & \frac{t_0}{U_l} e^{i(l+1)\gamma}
\end{eqnarray}
We observe that the ring-dot coupling $\lambda$ does not appear in any of these 
expressions. This is because of the selection $E=\tilde\epsilon_0$ and a non-zero $\lambda$, however 
small. Let us now define, $\cos(qa) = (E-\epsilon_0)/2t_0 = (\tilde\epsilon_0-\epsilon_0)/2t_0 = 
\delta/2t_0$, $\tilde\epsilon_0-\tilde\epsilon_L=\xi_1$ and $\tilde\epsilon_0-\tilde\epsilon_R=\xi_2$. 
With these, the transfer matrix elements for the {\it diatomic molecule} read, 
\begin{eqnarray}
M_{11} & = & \left [\frac{\xi_1\xi_2U_l}{t_0^2} - \frac{1}{U_l} \right ] e^{-i(l+1)\gamma} \nonumber \\
M_{12} & = & -\frac{\xi_2U_l}{t_0} e^{-i(l+1)\gamma} \nonumber \\
M_{21} & = & \frac{\xi_1U_l}{t_0} e^{-i(l+1)\gamma} \nonumber \\
M_{22} & = & -U_l e^{-i(l+1)\gamma}
\end{eqnarray}
Finally, the transmission coefficient is given by, 
\begin{equation}
T = \frac{4[1-\frac{\delta^2}{4t_0^2}]}{|d_1|^2 + |d_2|^2}
\end{equation}
where, 
\begin{eqnarray}
d_1 & = & \frac{e^{-i(l+1)\gamma}}{t_0} \left [\left ( \frac{\xi_1\xi_2U_l}{t_0^2} + 
\frac{U_l^2-1}{U_l} \right )\frac{\delta}{2} - (\xi_1+\xi_2)U_l \right ] \nonumber \\ 
d_2 & = & e^{-i(l+1)\gamma} \left (\frac{\xi_1\xi_2U_l}{t_0^2} - \frac{U_l^2+1}{U_l}  \right ) 
\sqrt{1-\frac{\delta^2}{4t_0^2}}
\end{eqnarray}
As we observe, $|d_1|^2$ and $|d_2|^2$ and hence $T$, are independent of the flux. That is, the AB-oscillations 
are suppressed whenever the Fermi energy coincides with any of the discrete 
eigenvalues of the isolated QD-array.

In view of the above calculation, a few pertinent observations should be given due importance. 
Let us detune $E$ slightly from an eigenvalue of the QD-array.  
That is, let's set $E-\tilde\epsilon_0=\Delta$, $\Delta$ being very small. How does the shape
 of the AB-oscillation get altered in the neighborhood of $E=\tilde\epsilon_0$ ? In this case we have,
\begin{equation}
\epsilon^* = \epsilon_0 + \frac{\lambda^2}{\Delta}
\end{equation}
Clearly, if $\lambda \ne 0$, but very very small so that $\lambda^2 \sim {\cal O}(\Delta)$, then the analysis
as given above, is not valid, as $\epsilon^*$ is not {\it infinity} any more. As a result, we shall
observe AB-oscillations in the transmission spectrum in general. If we gradually increase the value of 
$\lambda$ so that $\Delta \ll \lambda$, or even smaller, then we essentially keep on making 
$\lambda^2/\Delta$ and hence $\epsilon^{*}$ larger and larger. This results in the gradual suppression 
of the amplitude of the AB-oscillations, and 
finally, when $\lambda^2/\Delta$ becomes a very large number (dictated by the machine precision), the 
transmission coefficient ($T$) becomes independent of the flux threading the ring. AB-oscillations 
disappear completely.

An interesting feature of the AB oscillations in such cases is that, 
a gradual increase in the value of $\lambda$ can 
lead either to $T=1$ (Fig.2a) or to $T=0$ (Figs.2b and 2c). This depends on the combination of the 
size of the ring (i.e. on $l$, $m$ and $n$) and the length of the QD array ($N$). In every case however, 
the progress towards $T=1$ or $T=0$ is accompanied by a gradual suppression of the AB oscillations. 
Most interestingly, for a set of values of $l$, $m$, $n$ and $N$, the phase of the AB oscillations is 
reversed at specific values of the magnetic flux as soon as $\lambda$ exceeds some `critical' value. 
Incidentally, a similar observation in a simpler geometry was reported 
by Kubala and K\"{o}nig as well \cite{kub02}.
The present cases are depicted in Fig.$2$(b) and $2$(c) where the reversal is observed at $\Phi=\Phi_0/2$ and 
$3\Phi_0/2$ (odd multiple of $\Phi_0/2$ in general). However, with different combinations of $l$, $m$, 
$n$ and $N$, the reversal can take place at other flux values as well, for example at $\Phi=0$ and $\Phi=\Phi_0$.
We have not been able to obtain exact criterion for the phase reversal. However, extensive numerical search 
has revealed that this is true for various combinations of the size of the ring and the length of the QD array. 

Before ending, it should be mentioned that, the flux indepence that we have discussed above, is 
basically caused by the divergence of $\epsilon^{*}$ at special energies. This divergence can also 
be achieved for any arbitrary energy other that the eigenvalues of the 
isolated QD array, by letting $\lambda \rightarrow \infty$. Such a situation, as we have carefully observed, 
but do not report here to save space, leads to a flux independent $T-E$ spectrum.
\subsection{Selective switching}
\begin{figure}
{\centering \resizebox* 
{5cm}{3cm}{\includegraphics[angle=0]{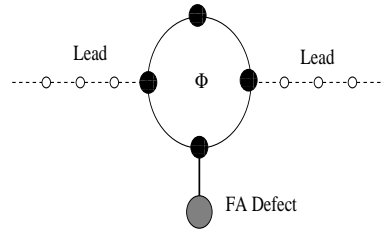}}\par}
\caption{\label{fig3} A simple $4$-site ring with a single FA defect attached to it}  
\end{figure}

At first, we note that, for $l=m+n+1$, i.e., for an equal number ($l$) atoms in the two arms, and 
with $\lambda=0$, the $L-R$ hopping integral in the effective {\it diatomic molecule} is real, and reads,
\begin{equation}
t_F = \frac{2t_0}{U_l} \cos(\frac{\pi\Phi}{\Phi_0})
\end{equation}
$t_B$ is of course, equal to $t_F$. It is now clear that for $\Phi=\Phi_0/2$, the effective hopping integral
becomes zero, resulting in $T=0$ (an antiresonance) independent of the energy of the electron. As soon as the 
ring-dot coupling $\lambda$ assumes a non-zero value, interesting transmission behavior is observed. To get 
a clearer understanding, we refer to Fig.3, which is a ring with just one atom in both the lower and the 
upper arms with a single QD with an on-site potential $\epsilon_D$ 
coupled to the atom in the lower arm. This is a simple modification of 
the model used by Kubala and K\"{o}nig \cite{kub02}. For this simple geometry, with $\lambda \ne 0$, we get, 
\begin{eqnarray}
\tilde\epsilon_L & = & \epsilon_0 + \frac{t_0^2}{E-\epsilon_0} + \frac{t_0^2(E-\epsilon_D)}
{(E-\epsilon_0)(E-\epsilon_D)-\lambda^2} \nonumber \\
t_F & = & (-i) t_0^2 \frac{\lambda^2}{(E-\epsilon_0)[(E-\epsilon_0)(E-\epsilon_D)-\lambda^2]}
\end{eqnarray}
with $\tilde\epsilon_R=\tilde\epsilon_L$ and $t_B=t_F^*$. Using these, one can work out the transfer matrix 
elements for the {\it diatomic molecule}, in the limit $E \rightarrow \epsilon_D$, to be equal to, 
\begin{eqnarray}
\lim_{E \rightarrow \epsilon_D} M_{11} & = & \frac{-i\delta(2t_0^2-\delta^2)}{t_0^3} \nonumber \\
\lim_{E \rightarrow \epsilon_D} M_{12} & = & \frac{-i(t_0^2-\delta^2)}{t_0^2} \nonumber \\
\lim_{E \rightarrow \epsilon_D} M_{21} & = & \frac{i\delta(2t_0^2-\delta^2)}{t_0^3} \nonumber \\
\lim_{E \rightarrow \epsilon_D} M_{22} & = & \frac{-i\delta}{t_0} 
\end{eqnarray}
where, $\delta=\epsilon_0-\epsilon_D$. Inserting these values in the formula for the transmission coefficient, 
it is observed that $T=1$ for $E=\epsilon_D$ with any $\lambda \ne 0$. That is, the presence of a finite 
ring-dot coupling, however small, triggers ballistic transmission across the ring. 

With an arbitrary number of scatteres in either arm of the ring, and the QD array extending beyond 
one atom, the situation is non-trivial and  
closed form expressions look extremely cumbersome to deal with. We have conducted extensive and 
careful numerical investigation to examine several cases. Here, details of a specific case are
given which reflect the generic features of the {\it selective switching} effect that we wish to 
highlight.

We choose a situation where the QD array contains, for example, 
 five atoms ($N=5$). The on-site potential $\epsilon_0$ 
and the hopping integral $t_0$ are set equal to zero and unity everywhere, including the QD array.
We set the magnetic flux $\Phi=\Phi_0/2$, and select $l=2m+1$ and $m=n$.
The $5$-site QD chain is now diagonalized to get the eigenvalues $0$, $\pm 1$ and $\pm \sqrt{3}$.
With $\lambda$ set equal to zero, as discussed before, we get 
$T=0$ irrespective of energy $E$. Interestingly, it is found that, by choosing a small non-zero value of 
$\lambda$ and an 
appropriate set of values for $l$, $m$ and $n$, but always satisfying the 
requirement $l=2m+1$ and $m=n$, it is possible to make the ring-dot system {\it completely 
transparent} to an incoming electron when its energy becomes equal to some or all of the eigenvalues of the 
$5$-site QD array. The transmission at any energy outside the set of five eigenvalues mentioned above can be 
completely suppressed if $\lambda$ is kept small enough. However, a gradual increase in the value of $\lambda$ 
gives rise to secondary transmission peaks as the ring `interacts' with the QD more strongly. 
These secondary peaks finally settle into bands of transmission separated by transmission dips, as a result 
of quantum interference.
The important 
thing to appreciate is that whether we observe complete transparency at a subset of the eigenvalues or 
for all of them, depends strongly on the mutual tuning of the values of the ring-dot coupling $\lambda$ and 
$l$, $m$ and $n$. Fig.4 displays the slective switching action when the QD array contains $3$ and $5$ sites 
respectively (Fig.4a and 4b). In Fig.4a, setting $\lambda=0.08$ and attaching the array to an $(l,m,n)=(1,0,0)$ 
ring (like one shown in Fig.2), we see that the transmission coefficient is unity (or very close to it) only 
when energy $E$ is equal to the three eigenvalues of the isolated $3$-dot array, viz, at
$E=0$ and $\pm\sqrt{2}$. On the other hand, with $N=5$ 
(Fig.4b), with $l=17$, $m=n=8$ and $\lambda=0.04$, transmission is triggered only at three of the five 
eigenvalues. The scenario of course changes as the parameters are varied, keeping $\lambda$ small. However, 
the `smallness' of $\lambda$ is to be selected by trial method, at least so far as we have checked.
In Table 1, we provide a list of such selective values for which $T=1$ (or very close to it) at $\Phi=\Phi_0/2$.
\begin{figure}
{\centering \resizebox* 
{16cm}{14cm}{\includegraphics[angle=-90]{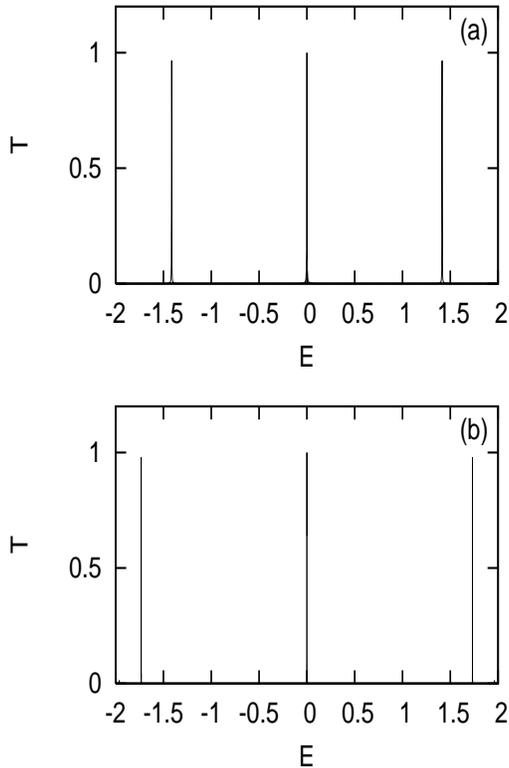}}\par}
\caption{\label{fig4} Selective switching effect at $\Phi=\Phi_0/2$ for 
a QD array of $3$ sites (a), and $5$ sites (b). In $4(a)$, we have taken  
$l=1$, $m=n=0$ and $\lambda=0.08$. Three transmission peaks at three distinct eigenvalues 
$E=0$ and $\pm \sqrt{2}$ of the isolated $3$-site QD chain are visible.
In (b) $l=17$, $m=n=8$, $N=5$ and $\lambda=0.04$. Peaks appear at $E=0, \pm \sqrt{3}$, 
and transmission at two other eigenvalues of the isolated QD chain, viz, at $E=\pm 1$ are blocked.}
\end{figure}
\vskip .3in
\noindent
\begin{tabular} {|l|c|r|} \hline
(l,m,n) & Typical value of $\lambda$ & $T\sim 1$ at $E=$ \\ \hline
(12k-1,6k-1,6k-1) & arbitrary & No peak at all \\
(12k-7,6k-4,6k-4) & 0.04 & $0$, $\pm \sqrt{3}$ \\
(12k-5,6k-3,6k-3) & 0.04-0.045 & $\pm 1$, $\pm \sqrt{3}$ \\ 
(12k+3,6k+1,6k+1) & 0.04-0.045 & $\pm 1$, $\pm \sqrt{3}$ \\ 
(12k-11,6k-6,6k-6) & 0.05-0.10 & $0$, $\pm 1$, $\pm \sqrt{3}$ \\ 
(12k-3,6k-2,6k-2) & 0.05-0.10 & $0$, $\pm 1$, $\pm \sqrt{3}$ \\ \hline
\end{tabular}
\vskip .25in
Table 1. Some typical combinations of $l$, $m$, $n$ and the ring-dot coupling $\lambda$ that 
give rise to selective swithing at $\Phi=\Phi_0/2$. $k$ is a positive integer.
\vskip .3in
Before we end, it should be noted that the geometry dealt with in the present communication can 
equivalently be thought as a discrete part (the lower arm plus the QD array) to an infinite linear
chain (the left lead plus the upper arm plus the right lead). Considering no magnetic field, we expect 
Fano lineshapes in the transmission spectrum as a result of an `interaction' of the discrete spectrum 
of the lower parts with the continuous spectrum offered by the upper section \cite{mir05},\cite{kiv05}. 
Indeed there are such lineshapes in the transmission resonances, which however get masked due to 
quantum interference as we take larger and larger size of the ring as well as the QD array. 
\vskip .3in
\section{conclusion}
We have addressed the issue of transmission across an Aharonov-Bohm ring with a dangling chain of 
single level quantum dots within a tight binding formalism. In presence of a magnetic flux threading the 
ring, we discuss the role of the ring-dot coupling in controlling the profile of transmission oscillations.
The central feature is a suppression of the AB oscillations with occasional reversal of phase at specific values 
of the flux. 
Most interestingly, it is found that, a simultaneous adjustment of the number of scatterers in the arms of 
the ring and 
the ring-dot coupling can lead to a complete transparency of the system at some or all of the eigenvalues 
of the QD array. It is important to note that a bigger ring with large values of $l$, $m$, $n$ (always 
satisfying the condition $m=n$, $l=2m+1$ and $\Phi=\Phi_0/2$) exhibits ballistic transmission $T=1$ for 
rather low values of the ring-dot coupling $\lambda$. This is because, with a bigger ring the coupled 
QD array stays far away from the junctions $L$ and $R$. The `end effects' are thus minimised. We have also 
tested these features with a QD array formed according to the quasiperiodic Fibonacci growth rule \cite{arun06}.
The essential features like the self-similarity in the electronic transmission are also observed in the 
selective switching case. Such aspects will be discussed elsewhere.
\vskip .3in
\noindent
{\bf References}
\vskip .25in
\begin{itemize}
\bibitem{kub02} B. Kubala and J. K\"{o}nig, Phys. Rev. B {\bf 65}, 245301 (2002). 
\bibitem{zeng02} Z. Y. Zeng, F. Claro, and A. P\'{e}rez, phys. Rev. B {\bf 65}, 085308 (2002).
\bibitem{torio02} M. E. Torio, K. Hallberg, A. H. Cecatto, and C. R. Pretto, Phys. Rev. B {\bf 65},
085302 (2002).
\bibitem{pou02} V. Pouthier and G. Girardet, Phys. Rev. B {\bf 66}, 115322 (2002). 
\bibitem{lad03}  M. L. Ladr\'{o}n de Guevara, F. Claro, and P. A. Orellana, Phys. Rev. B {\bf 67}, 195335 (2003).
\bibitem{ore03} P. A. Orellana, F. Dom\'{i}nuez-Adame, I. G\'{o}mez, and M. L. Ladr\'{o}n de Guevara, 
Phys. Rev. B {\bf 67}, 085321 (2003).  
\bibitem{rod03} A. Rodr\'{i}guez, F. Dom\'{i}nguez-Adame, I. G\'{o}mez, and P. A. Orellana, 
Phys. Lett. A {\bf 320}, 242 (2003). 
\bibitem{xiong04} Y. -J. Xiong and X. -T. Liang, Phys. Lett. A {\bf 330}, 307 (2004).
\bibitem{paore04} P. A. Orellana, M. L. Ladr\'{o}n de Guevara, and F. Claro, Phys. Rev. B {\bf 70},
233315 (2004).
\bibitem{gom04} I. G\'{o}mez, F. Dom\'{i}nguez-Adame, and P. Orellana, J. Phys.:Condens. Matter 
{\bf 16}, 1613 (2004). 
Phys. Rev. B {\bf 70}, 035319 (2004).
\bibitem{torio04} M. E. Torio, K. Hallberg, A. E. Miroshnichenko, and M. Titov, Eur. Phys. J. B {\bf 37}, 
399 (2004).  
\bibitem{adam04} F. Dom\'{i}nguez-Adame, I. G\'{o}mez, P. A. Orellana, and M. L. Ladr\'{o}n 
de Guevara, Microelectronics. Jour. {\bf 35} 87 (2004).
\bibitem{mir05} A. E. Miroshnichenko and Y. S. Kivshar, Phys. Rev. E {\bf 72}, 056611 (2005). 
\bibitem{arun06} A. Chakrabarti, Phys. Rev. B {\bf 74}, 205315 (2006).
\bibitem{arun07} A. Chakrabarti, Phys. Lett. A {\bf 366} 507 (2007).
\bibitem{kiv05} A. E. Miroshnichenko, S. F. Mingaleev, S. Flach, and Y. S. Kivshar, 
Phys. Rev. E {bf 71}, 036626 (2005).
\bibitem{bao05} K. Bao and Y. zheng, Phys. Rev. B {\bf 73}, 045306 (2005).
\bibitem{li05} H. Li, T. L\"{u}, and P. Sun, Phys. Lett. A {\bf 343}, 403 (2005).
\bibitem{mar05} M. Mardaani and K. Esfarjani, Physica E {\bf 27}, 227 (2005).
\bibitem{xiong06} Z.-B. He and Y.-J. Xiong, Phys. Lett. A {\bf 349}, 276 (2006). 
\bibitem{ab59} Y. Aharonov and D. Bohm, Phys. Rev. {\bf 115}, 485 (1959).
\bibitem{butt83} M. B\"{u}ttiker, Y. Imry, and R. Landauer, Phys. Lett. A {\bf 96}, 365 (1983).
\bibitem{gefen84} Y. Gefen, Y. Imry, and M. Ya. Azbel, Phys. Rev. Lett. {\bf 52}, 129 (1984).
\bibitem{amato89} J. L. D'Amatao, H. M. Pastawski, and J. F. Weisz, Phys. Rev. B {\bf 39}, 3554 (1989).
\bibitem{aldea92} A. Aldea, P. Gartner, and I. Corcotoi, Phys. Rev. B {\bf 45}, 14122 (1992).
\bibitem{yacob95} A. Yacoby, M. Heiblum, D. Mahalu, and H. Shtrikman, Phys. Rev. Lett. {\bf 74}, 
4047 (1995).
\bibitem{schus97} R. Schuster, E. Buks, M. Heiblum, D. Mahalu, V. Umansky, and H. Shtrikman, 
Nature (London) {\bf 385}, 417 (1997).
\bibitem{holl01} A. W. Holleitner, C. R. Decker, H. Qin, K. Eberl, and R. H. Blick, 
Phys. Rev. Lett. {\bf 87}, 256802 (2001).
\bibitem{levy95} A. Levy Yeyati and M. B\"{u}ttiker, Phys. Rev. B {\bf 52}, R14360 (1995).
\bibitem{kang99} K. Kang, Phys. Rev. B {\bf 59}, 4608 (1999).
\bibitem{meier04} L. Meier, A. Fuhrer, T. Ihn, K. Ensslin, W. Wegscheider, and M. Bichler, 
Phys. Rev. B {\bf 69}, 241302(R) (2004).
\bibitem{koba02} K. Kobayashi, H. Aikawa, S. Katsumoto, and Y. Iye, Phys. Rev. Lett. {\bf 88}, 
256806 (2002).
\bibitem{koba04} K. Kobayashi, H. Aikawa, A. Sano, S. Katsumoto, and Y. Iye, 
Phys. Rev. B {\bf 70}, 035319 (2004).
\bibitem{lee06} M. Lee and C. Bruder, Phys. Rev. B {\bf 73}, 085315 (2006); R. Wang and J.-Q. Liang, 
Phys. Rev. B {\bf 74}, 144302 (2006).
\bibitem{fano61} U. Fano, Phys. Rev. {\bf 124}, 1866 (1961).
\bibitem{grim00} X. Wang, U. Grimm, and M. Schreiber, Phys. Rev. B {\bf 62}, 14020 (2000);
J. Q. You and Q. B. Yang, J. Phys.:Condens. Matter {\bf 2}, 2093 (1990).
\bibitem{stone81} A. Douglas Stone, J. D. Joannopoulos, and D. J. Chadi, Phys. Rev. B {\bf 24}, 5583 (1981).
\end{itemize}
\end{document}